\documentclass[12pt]{article}
\usepackage[top=2cm,bottom=2cm,left=1.5cm,right=1.5cm,columnsep=0.5cm]{geometry}
\usepackage{graphicx}
\usepackage{amsmath,amssymb}
\usepackage{bm}
\usepackage{hyperref}
\usepackage{booktabs}
\usepackage{xcolor}
\usepackage{microtype}
\usepackage{float}
\usepackage[figuresleft]{rotating} 

\usepackage{abstract}
\usepackage{tikz}
\usetikzlibrary{positioning,shapes.geometric,decorations.pathreplacing}

\begin{document}

\title{The Pauli Lightcone: Information-Theoretic Error Mitigation\\
       Beyond the Autocorrelation}

\author{
 Paolo D'Alberto\thanks{Advanced Micro Devices, Inc. (AMD).
 \texttt{paolo.dalberto@amd.com}.
 This work was developed in active collaboration with Claude (Anthropic).}\\
 \small Advanced Micro Devices, Inc.\\
 \small\texttt{paolo.dalberto@amd.com}
}

\date{\today}

\maketitle

\begin{abstract}
We simulate the kicked Heisenberg Ising model on heavy-hex and
rectangular lattices using a GPU-accelerated tensor network simulator
and study the non-identity Pauli weight $n(v,t)$, the probability
that the evolved operator has support at site $v$ at cycle $t$, as a
space-time observable.  Under noise amplification at levels
$\gamma=1,2,3$, $n(v,t)$ directly reveals how noise deforms the
operator lightcone across the lattice.

We introduce the \emph{wavemap}: a spatial portrait of noise effects
that assigns each site a per-noise-level arrival delay $l_\gamma(v)$
and cross-entropy loss $\mathcal{L}_\gamma(v)$.  These observables are
exact at the lightcone frontier, where bond dimension $\chi\approx1$
and the simulation is most faithful.  Eigenvalue analysis of the
composed gate-plus-noise Pauli transfer matrices confirms that the
studied noise is pure amplitude damping: the spatial propagation
pattern is entirely determined by the gate, making the wavemap a
model-free noise diagnostic.

We apply the multi-product formula (MPF) to recover the noiseless
Pauli weight field from the noisy samples, subject to the
Lieb-Robinson causal constraint $n_{\rm MPF} \leq n_{\rm nl}$.
Fitting time-adaptive coefficients $\alpha(t)$ over the frontier
recovers up to $\sim$55\% of the information loss relative to the
best noisy sample, exploiting the fact that the frontier is where
truncation error is smallest.  On a heavy-hex lattice with uniform synthetic noise, the method
recovers 49\% of information loss despite only 21/68 sites inside
the noiseless lightcone.
\end{abstract}

\section{Introduction}
\label{sec:intro}

\begin{figure}[t]
\centering
\includegraphics[width=\columnwidth]{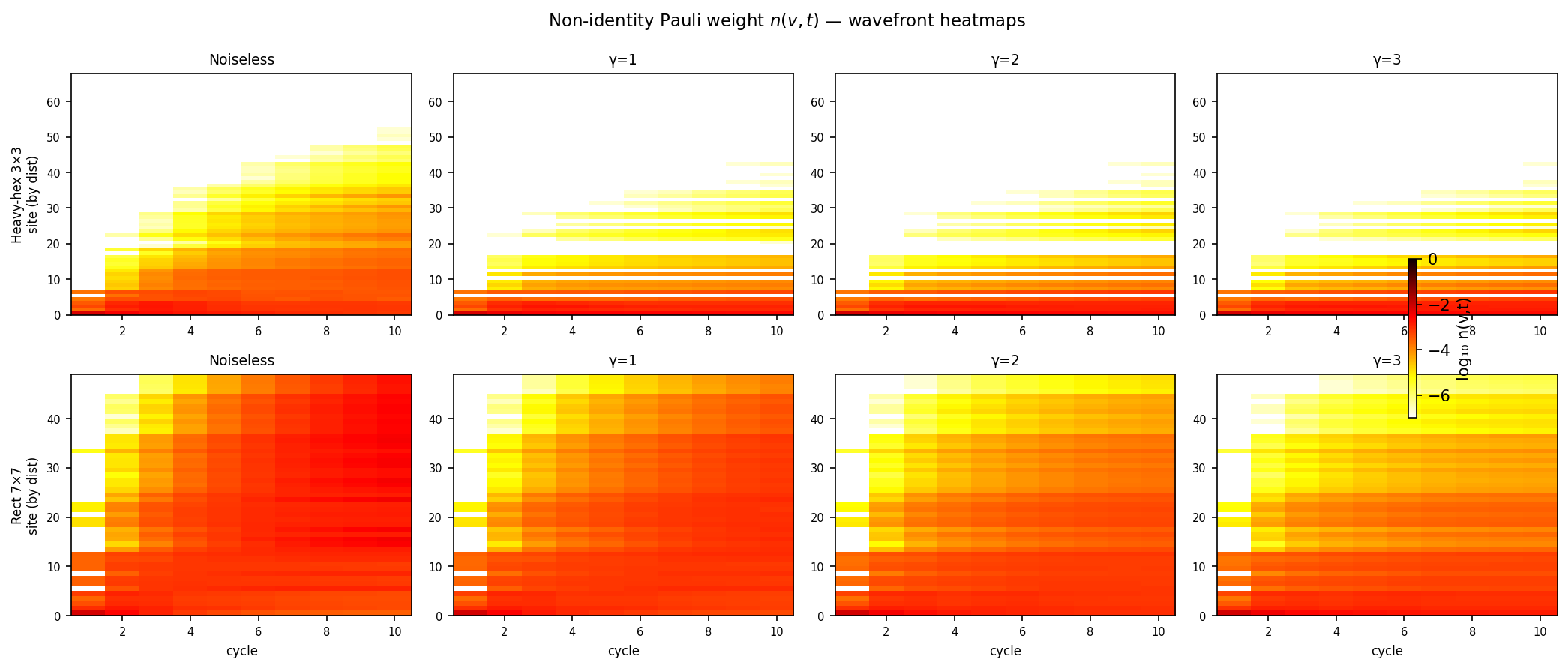}
\caption{Non-identity Pauli weight $n(v,t)$ per site per cycle for noiseless (NL)
and three noise levels ($\gamma=1,2,3$) on the heavy-hex $3\times3$ lattice
(top) and rect $7\times7$ (bottom), both initialized at the center site.
Color encodes $\log_{10} n(v,t)$; white = below threshold.
The wavefront advances diagonally in the (space, time) plane at velocity $v_{\rm hop}$.
Noise suppresses amplitudes, pushing the threshold-crossing later, a uniform
shift of the wavefront, not a change of its spatial pattern.}
\label{fig:heatmap}
\end{figure}

When a local operator evolves under a noisy Hamiltonian, its Pauli weight
$n(v,t)$ spreads outward from the initialization site as a wave.
Compared to the noiseless reference, the noisy wave travels slower, reaches
fewer sites, and arrives with less amplitude.
This is not a model prediction; it is a direct observation from the data,
visible without fitting or assumptions, see Figure~\ref{fig:heatmap}.
A different noise regime, with off-diagonal Pauli transfer matrix terms,
can inject weight at unvisited sites, making parts of the frontier appear
faster rather than slower, a qualitatively distinct signature.

This observation motivates the central question of this paper:
can we quantify, site by site and cycle by cycle, how much information
the noisy evolution has lost relative to the noiseless reference?

The standard MPF approach~\cite{mpf-ref} fits the autocorrelation
$C(t) = \langle Z_c(t) Z_c(0)\rangle$ across noise levels $\gamma$
and extrapolates to zero noise.
This is a single scalar per cycle, compressing the entire space-time evolution
into one number at each $t$.
Two limitations follow.
First, spatial information is lost: sites strongly affected by noise and sites
barely affected both contribute to the same scalar $C(t)$.
Second, $C(t)$ is dominated by contributions from the interior of the
lightcone, where bond dimension $\chi$ is large and belief propagation is
approximate. The wavefront, where $\chi\approx1$ and the simulation is exact,
contributes least to $C(t)$.

The non-identity Pauli weight $n(v,t)$ resolves both limitations.
Any operator $O_c(t)$ can be expanded in the $n$-qubit Pauli basis
$\{P\} = \{I,X,Y,Z\}^{\otimes n}$ as $O_c(t) = \sum_P c_P(t)\, P$,
where the real coefficients $c_P(t)$ satisfy $\sum_P c_P(t)^2 = 1$.
Then $n(v,t)$ is defined as the total weight on Paulis that act
non-trivially at site $v$:
\begin{equation}
    n(v,t) = \sum_{\substack{P:\, P_v \neq I}} c_P(t)^2
    \label{eq:nvt}
\end{equation}
By normalization, $n(v,t) \in [0,1]$ is the probability that a Pauli
string sampled from the operator distribution acts non-trivially at site $v$~\cite{nahum2018operator,Tan2025Operator}.
It is a probability distribution over the lattice, evolving in time as the
operator spreads. At the wavefront it is computed exactly at any $\chi$:
frontier sites carry only nascent correlations and $\chi\approx1$ by
construction. This is where the simulation is most faithful and where noise
effects are most cleanly measurable.

The \emph{frontier} at cycle $t$ is the set of sites that first acquire
significant operator weight at that cycle:
\begin{equation}
    \mathcal{F}(t) = \bigl\{ v : n(v,t) > \varepsilon_{\rm th}
                             \text{ and } n(v,t-1) \leq \varepsilon_{\rm th} \bigr\}
    \label{eq:frontier}
\end{equation}
The frontier is the leading edge of the lightcone, where $\chi\approx1$ and
the simulation is exact. For each site $v$ and noise level $\gamma$, we
define the \emph{arrival delay} $l_\gamma(v) = t_\gamma(v) - t_{\rm nl}(v)$
as the number of cycles by which the noisy frontier lags the noiseless one,
and the \emph{cross-entropy loss}
\begin{equation}
    \mathcal{L}_\gamma(v) = \sum_{t} -\ln\frac{n_\gamma(v,t)}{n_{\rm nl}(v,t)}
    \label{eq:loss}
\end{equation}
summed over a delay-adaptive window centered on arrival, as the accumulated
information loss at site $v$.
Together, $l_\gamma(v)$ and $\mathcal{L}_\gamma(v)$ form the \emph{wavemap},
a spatial portrait of noise effects across the lattice, and the
space-time trajectory of the frontier $\mathcal{F}(t)$ connects the
per-site observables to the global propagation dynamics.

Several lines of prior work inform this paper.
Operator spreading in random unitary circuits~\cite{nahum2018operator,Tan2025Operator}
established the butterfly velocity and domain-wall width as natural observables
for wavefront characterization; our work applies analogous ideas to structured
Hamiltonian evolution with noise amplification.
Real-time operator evolution in two and three dimensions via sparse Pauli
dynamics~\cite{Begusic2025RealTime} and Pauli propagation under arbitrary local
noise~\cite{rudolph2025simulating} provide the simulation framework on which
CppSim is based.
Noise characterization from circuit measurements using sparse Pauli-Lindblad
models~\cite{Hu2025Demonstration} and self-consistent Pauli noise
learning~\cite{Chen2026Efficient} are complementary approaches that infer noise
rates by fitting a model; our approach uses the spatial wavefront pattern to
characterize noise without model assumptions.
The multi-product formula for error mitigation was first applied to the
scalar autocorrelation $C(t)$ in tensor network simulations of the heavy-hex
lattice~\cite{mpf-ref}, and recently extended to operator inner products
$\langle\psi_{\gamma_i}|\psi_{\gamma_j}\rangle$ on IBM quantum
hardware~\cite{robertson2026lindbladmultiproductformulas,dalberto2026cppsim}.
Our work takes a different direction: rather than autocorrelations or inner
products, we use the non-identity Pauli weight field $n(v,t)$ — the diagonal
of the operator's Pauli density matrix — as a site-resolved space-time
observable, with cross-entropy loss as the MPF objective.

We show via PTM eigenvalue analysis that the noise models studied here are
pure amplitude damping, leaving the spatial propagation pattern determined
by the gate.
The Lieb-Robinson bound~\cite{liebrobinson1972} states that information
cannot propagate faster than a finite velocity in any local lattice Hamiltonian,
implying $n_\gamma(v,t) \leq n_{\rm nl}(v,t)$ for pure damping noise:
the noiseless evolution is the causal ceiling.
Entropy-optimal MPF with this constraint $n_{\rm MPF} \leq n_{\rm nl}$
recovers $\sim$25\% more information than the best noisy sample on a uniform-noise
rectangular lattice.
On a heavy-hex lattice with uniform synthetic noise, entropy-optimal MPF
recovers 49\% of information loss, demonstrating that sparse lightcone
coverage alone does not limit recovery.

\section{Hamiltonian, Noise, and Eigenvalue Analysis}
\label{sec:setup}

We simulate the kicked isotropic Heisenberg model in the Heisenberg picture:
\begin{equation}
    H = J \sum_{\langle ij\rangle} \bigl(X_iX_j + Y_iY_j + Z_iZ_j\bigr),
    \qquad
    U_{\rm kick} = \prod_v R_x(2\phi)_v,\quad \phi = 1.41371669
\end{equation}
with uniform per-bond coupling $J_{\langle ij\rangle} = 1$ and Trotter step
$\varepsilon = 0.10$ for all results presented here.
The coupling constants are simulation parameters that can be replaced by
device-characterized or otherwise physically motivated values without
modifying the wavemap methodology.
Each Trotter cycle applies the bond gates across all color classes in sequence,
followed by a single kick gate $R_x(2\phi)$ at every site.
We use CppSim~\cite{dalberto2026cppsim}, a GPU-accelerated belief propagation
tensor network simulator in which operators are evolved via $16\times16$ real
Pauli transfer matrices (PTMs), with bond dimension $\chi$ controlled by
Householder QR truncation at cutoff $\delta$.
All results use $\varepsilon=0.10$, $\delta=10^{-3}$, output precision $10^{-7}$,
and threshold $\varepsilon_{\rm th}=10^{-4}$.

Two implementation choices are worth stating explicitly.
First, the sub-gates within each bond ($G_1\ldots G_4$ and interleaved noise
channels) are pre-composed into a single $16\times16$ PTM per bond before
simulation; belief propagation and truncation are applied once per full Trotter
cycle, after all color classes have been applied.
Belief propagation is applied once per full Trotter cycle, after all color
classes, making the computation independent of color ordering.
Second, $n(v,t)$ is recorded once per full Trotter cycle, after the kick gate
fires; it is a snapshot of the operator state at the end of each complete cycle.
The wavefront and all wavemap observables are defined in terms of these
end-of-cycle snapshots.
A hardware experiment or a simulation that measures mid-cycle would observe
a different wavefront trajectory; our observables are specific to the
post-kick measurement point.

We study two topologies under two noise models.
For the heavy-hex $3\times3$ lattice (68 sites, 76 bonds), per-bond
noise PTMs representative of a physical heavy-hex device are interleaved
with the gate decomposition (uniform $J=1$ couplings; noise PTMs can be
replaced by device-characterized values without modifying the method):
\begin{equation}
    \mathrm{PTM}_{\rm noisy}^{(b)} =
    G_1 \cdot M_\gamma^{(b)} \cdot G_2(\alpha_b) \cdot M_\gamma^{(b)}
    \cdot G_3(\alpha_b) \cdot M_\gamma^{(b)} \cdot G_4(\delta_b)
    \label{eq:noisy_ptm}
\end{equation}
where $M_\gamma^{(b)} = \gamma M_{\rm hw}^{(b)} + (1-\gamma)I$ amplifies
the hardware noise channel by factor $\gamma$, and $G_1\ldots G_4$ implement
the hardware gate decomposition.
For the rectangular $7\times7$ lattice (49 sites, 84 bonds), we use synthetic
depolarizing noise: rows $1\ldots15$ of the ideal bond PTM $G^{(b)}$ are
scaled by $(1-p\gamma)$ with $p=0.01$, uniformly damping all non-identity
output Pauli components.

Before studying the wavemap observables, we establish the noise character
directly from the composed PTMs.

\begin{table}[H]
\centering
\small
\begin{tabular}{l r rr r r}
\toprule
Configuration & Bonds & Off-diag frac. & $\max|\mathrm{Im}(\lambda)|$
              & $n(\mathrm{Re}<0)$ & $n(|\lambda|>1)$ \\
\midrule
Heavy-hex ideal      & 76 & 0.357 & 0.964 & 0 & 0 \\
Heavy-hex $\gamma=1$ & 76 & 0.356 & 0.950 & 0 & 0 \\
Heavy-hex $\gamma=2$ & 76 & 0.355 & 0.950 & 2 & 0 \\
Heavy-hex $\gamma=3$ & 76 & 0.355 & 0.950 & 2 & 0 \\
\midrule
Rect $7\times7$ ideal      & 84 & 0.360 & 0.891 & 0 & 0 \\
Rect $7\times7$ $\gamma=1$ & 84 & 0.360 & 0.882 & 0 & 0 \\
Rect $7\times7$ $\gamma=2$ & 84 & 0.360 & 0.873 & 0 & 0 \\
Rect $7\times7$ $\gamma=3$ & 84 & 0.359 & 0.865 & 0 & 0 \\
\bottomrule
\end{tabular}
\caption{Eigenvalue statistics of composed noisy PTMs across all bonds.
Off-diagonal fraction = off-diagonal power / total power (averaged over bonds).
The fraction is constant across $\gamma$ ($\Delta < 0.002$) in both topologies,
confirming that Pauli mixing is entirely from the gate.
The 2 negative eigenvalues at $\gamma=2,3$ in heavy-hex are at $\mathrm{Re}=-10^{-5}$,
numerical precision rather than a structural change; they are nonetheless
relevant for the symmetric gauge, where $\sqrt{\lambda}$ must be computed to
absorb BP messages into site tensors: a negative eigenvalue makes this square
root imaginary and requires explicit sign handling in float32 arithmetic~\cite{dalberto2026cppsim}.}
\label{tab:eig}
\end{table}

\begin{figure}[H]
\centering
\includegraphics[width=0.65\columnwidth]{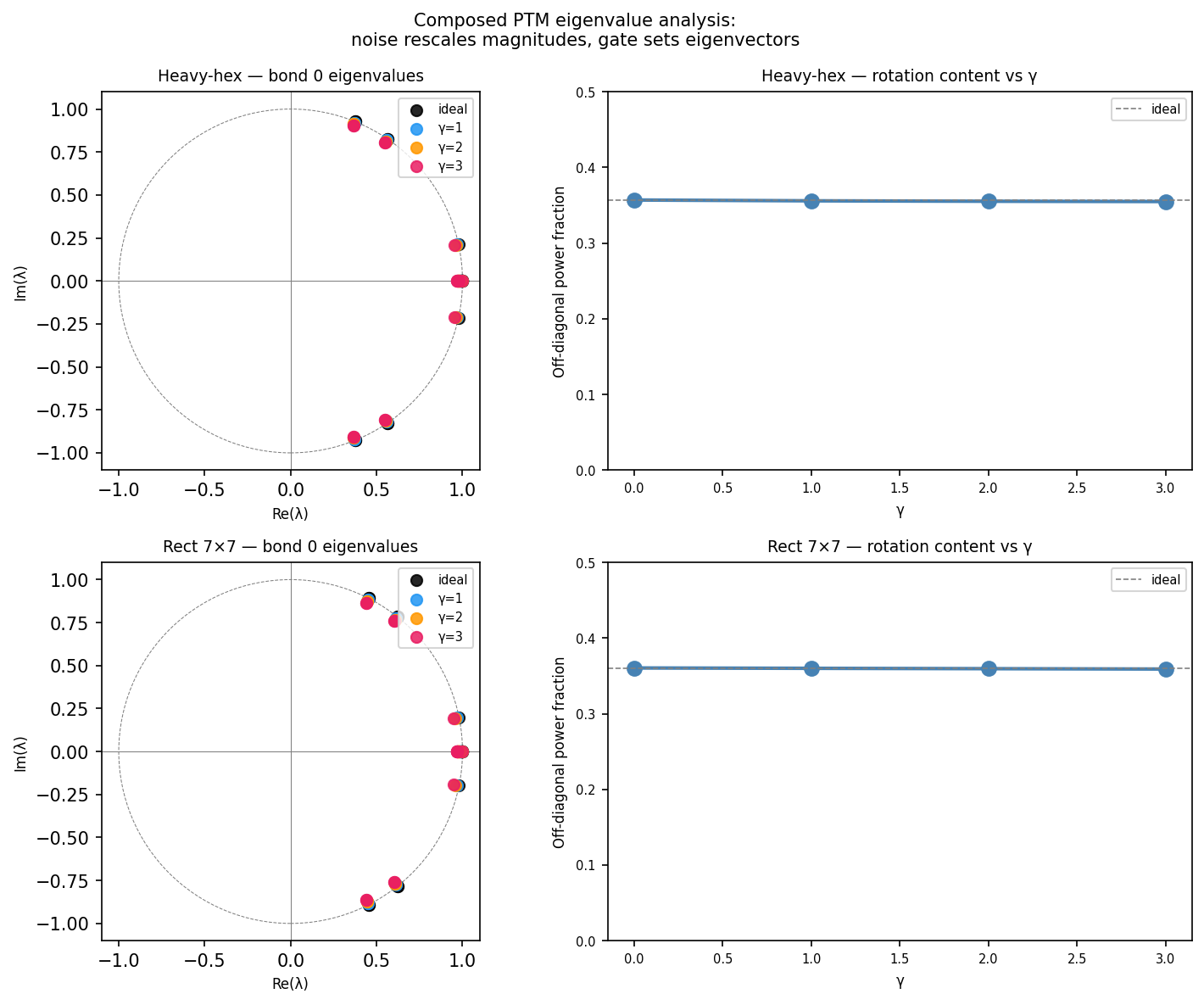}
\caption{Eigenvalue spectra of the composed noisy PTMs for one representative
heavy-hex bond (top) and one rect $7\times7$ bond (bottom), across $\gamma=1,2,3$.
Eigenvalue positions are nearly unchanged across $\gamma$: noise rescales magnitudes
slightly without rotating eigenvectors.
The off-diagonal power fraction (36\%) is constant; Pauli mixing is a gate
property, not a noise property.}
\label{fig:eig}
\end{figure}

Table~\ref{tab:eig} and Figure~\ref{fig:eig} establish the key result: the off-diagonal fraction is
constant across $\gamma$ ($\Delta < 0.002$) in both topologies.
Noise adds no rotation structure; it only scales eigenvalue magnitudes,
leaving all Pauli mixing to the gate $G$.
The noise channel is therefore a pure damping operator at the PTM level:
it delays the wavefront by reducing amplitudes below threshold without
redirecting the spatial propagation pattern, which remains entirely
determined by the Hamiltonian.

\section{The Pauli Lightcone and Wavefront}
\label{sec:observables}

For each site $v$ and noise level $\gamma$, the \emph{arrival time}
$t_\gamma(v) = \min\{t : n_\gamma(v,t) > \varepsilon_{\rm th}\}$
is the first cycle at which the wave crosses threshold,
and the \emph{arrival delay}
\begin{equation}
    l_\gamma(v) = t_\gamma(v) - t_{\rm nl}(v)
    \label{eq:delay}
\end{equation}
is the number of cycles by which the noisy wave lags the noiseless one.
For the noise models studied here (pure amplitude damping, Section~\ref{sec:setup}),
$n_\gamma(v,t) \leq n_{\rm nl}(v,t)$ at every site and cycle, so $l_\gamma(v) \geq 0$ always.%
\footnote{Rotation-type noise, with off-diagonal PTM terms injecting weight at
unvisited sites, can make the noisy wave arrive \emph{before} the noiseless at
isolated sites, giving $l_\gamma(v) < 0$. The Lieb-Robinson bound still holds
for the noiseless evolution, bounding its propagation speed; what can be inverted
is the site-by-site ordering between noisy and noiseless, since rotation noise
injects weight independently of the causal structure of the Hamiltonian.
In that regime the wavemap requires a signed delay.}

The ratio $r_\gamma(v) = n_\gamma(v, t_\gamma(v)) / n_{\rm nl}(v, t_\gamma(v))$
compares both waves at the moment of noisy arrival.
At that cycle the noisy wave has just crossed threshold
($n_\gamma \approx \varepsilon_{\rm th}$), while the noiseless wave has been
growing at $v$ for $l_\gamma(v)$ extra cycles
($n_{\rm nl} \gg \varepsilon_{\rm th}$).
The ratio $r_\gamma(v) \leq 1$ measures the combined effect of cumulative
path-dependent damping and the head start the noiseless wave gained during
the delay. It is bounded: $r_\gamma(v) \to 1$ when both waves arrive
simultaneously ($l_\gamma(v) = 0$) and decreases as the delay grows.

The ratio $r_\gamma(v)$ is a first step toward a probability comparison:
since $n(v,t)$ is a probability (Eq.~\ref{eq:nvt}), the ratio
$n_\gamma(v,t)/n_{\rm nl}(v,t)$ measures how much of the noiseless
probability at site $v$ and cycle $t$ the noisy evolution retains.
The natural information-theoretic measure of this loss is the
cross-entropy, which we sum over a delay-adaptive window
$[t_\gamma(v) - \Delta,\, t_\gamma(v) + \Delta]$ with
$\Delta = \max(1, l_\gamma(v))$:
\begin{equation}
    \mathcal{L}_\gamma(v) = \sum_{t \in \text{window}} -\ln\frac{n_\gamma(v,t)}{n_{\rm nl}(v,t)}
    \label{eq:loss2}
\end{equation}
This is the accumulated information loss at site $v$: zero when the two
distributions coincide, positive when noise suppresses the probability,
and with window width set by the delay itself so no free parameter is introduced.

\paragraph{Connection to hardware experiment.}
The noise amplification parameter $\gamma$ is not a simulation artifact:
it maps directly to the gate-folding factor used in zero-noise extrapolation
(ZNE)~\cite{giurgica2020digital}.
On real hardware, running the same circuit at fold factors $\gamma=1,2,3$
multiplies the effective noise per gate by $\gamma$ while leaving the
Hamiltonian structure unchanged, exactly the $M_\gamma$ model of
Eq.~\eqref{eq:noisy_ptm}.
The non-identity Pauli weight $n(v,t)$ is directly measurable per site per
cycle via local Pauli expectation values, accessible through randomized
benchmarking or Pauli noise tomography~\cite{rudolph2025simulating}.
The wavemap observables $l_\gamma(v)$ and $\mathcal{L}_\gamma(v)$ then
follow from the same threshold-crossing and cross-entropy computation
applied to the experimental $n(v,t)$ data; no simulation required.
The present work provides the simulation ground truth and the fitting
methodology; hardware validation is a direct next step.

\begin{figure}[p]
\centering
\includegraphics[width=\columnwidth]{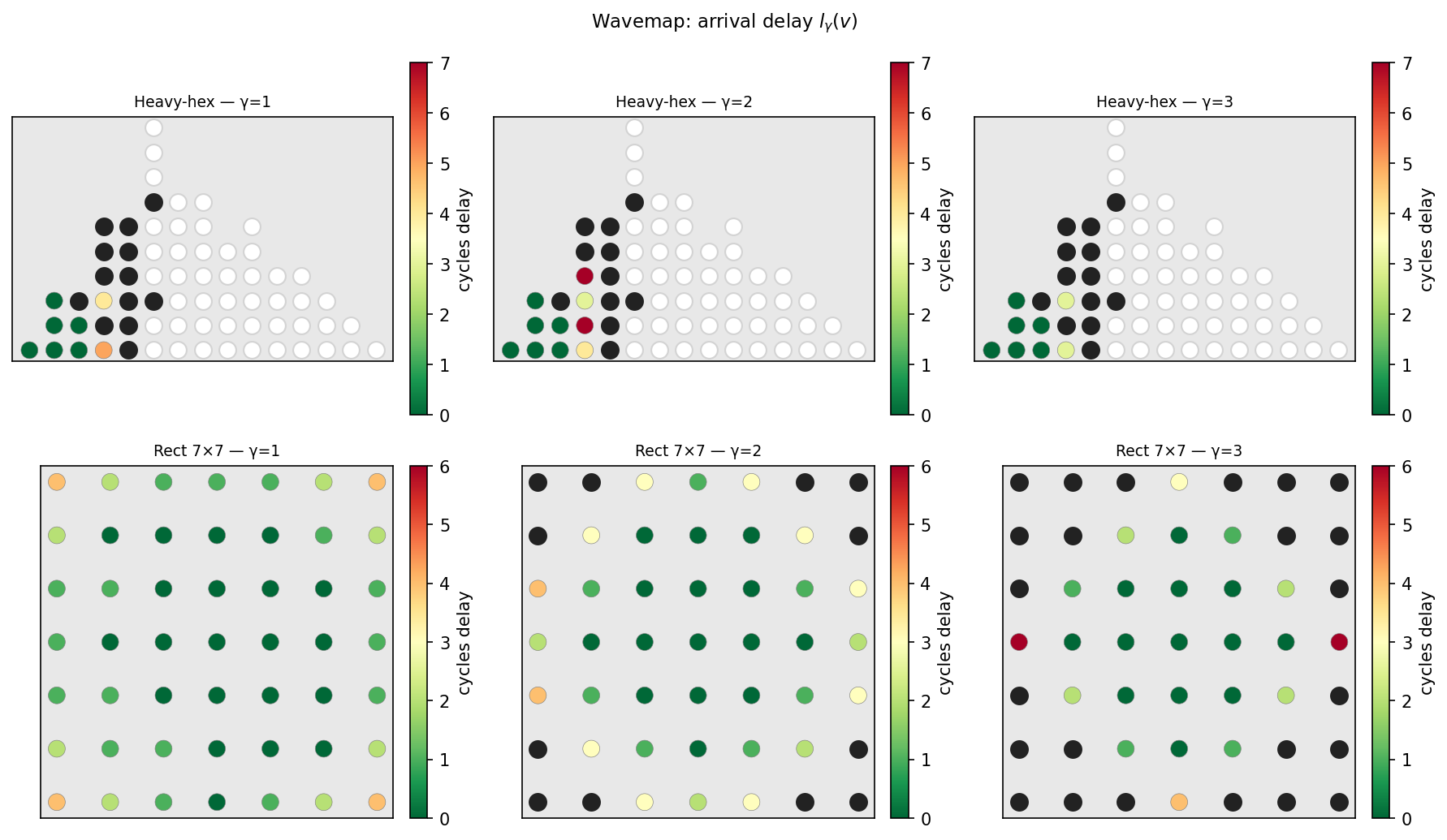}\\[1em]
\includegraphics[width=\columnwidth]{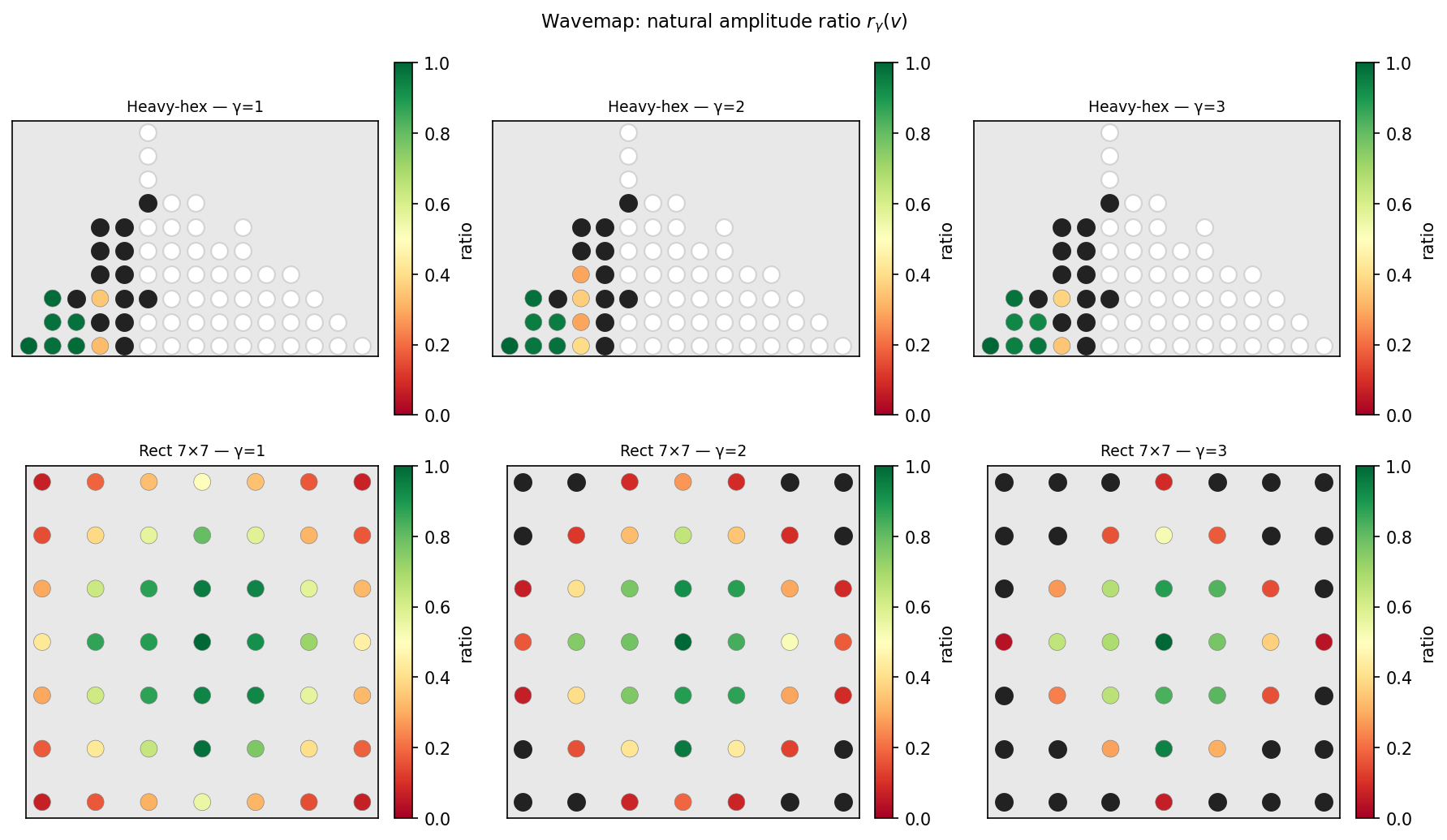}
\caption{Wavemap for rect $7\times7$ and heavy-hex, $\gamma=1,2,3$.
\textbf{Top}: arrival delay $l_\gamma(v)$ in cycles (white = outside lightcone,
black = noiseless arrived but noisy did not).
\textbf{Bottom}: amplitude ratio $r_\gamma(v) = n_\gamma(v,t_\gamma)/n_{\rm nl}(v,t_\gamma)$;
all values $\leq 1$ confirm the Lieb-Robinson bound.}
\label{fig:wavemap}
\end{figure}

The delay field $l_\gamma(v)$, the ratio $r_\gamma(v)$, the entropy loss
(Figure~\ref{fig:wavemap}),
$\mathcal{L}_\gamma(v)$, and the space-time frontier $\mathcal{F}(t)$
together form the \emph{wavemap}: a spatial portrait of how noise deforms
the Pauli lightcone, site by site.
All wavemap observables are exact at any bond dimension $\chi$, since frontier
sites have $\chi\approx 1$ by construction, and all are causally bounded
by the Lieb-Robinson structure.
In the following section we show how these observables can be refined through
the multi-product formula: fitting the wavemap across noise levels $\gamma$
and extrapolating toward the noiseless reference, with the Lieb-Robinson bound
as a causal constraint, yields a spatially-resolved error correction of the
operator lightcone.

\newpage
\section{Results}
\label{sec:results}

We present results in three stages, each building on the previous.
First, we characterize the global wavefront: the frontier velocity $v_{\rm hop}$
and per-site delay distribution $l_\gamma(v)$ establish how noise slows the
wave and which sites it fails to reach.
Second, we introduce the cross-entropy loss $\mathcal{L}_\gamma(v)$ as a
per-site information measure and show how its spatial and per-shell distributions
reveal the structure of noise across the lattice.
Third, we apply the multi-product formula to the wavemap in two forms:
a frontier-position MPF that recovers the noiseless wavefront trajectory,
and an entropy-optimal MPF that directly minimizes information loss subject
to the Lieb-Robinson causal constraint, demonstrating measurable information
recovery on the uniform-noise topology.

\subsection{Frontier velocity}
\label{sec:vhop}

\begin{table}[h]
\centering
\small
\begin{tabular}{l rrrr}
\toprule
Topology & $v_{\rm nl}$ & $v_{\gamma=1}$ & $v_{\gamma=2}$ & $v_{\gamma=3}$ \\
\midrule
Heavy-hex $3\times3$ ($\chi=200$) & 0.376 & 0.224 & 0.127 & 0.127 \\
Rect $7\times7$      ($\chi=50$)  & 0.448 & 0.382 & 0.248 & 0.097 \\
\bottomrule
\end{tabular}
\caption{Frontier velocity $v_{\rm hop}$ (hops/cycle), cutoff $\delta=10^{-3}$, 10 cycles,
uniform synthetic depolarizing noise on both topologies.
Rect $7\times7$: strictly monotone $v_{\rm nl}>v_1>v_2>v_3$.
Heavy-hex: near-degeneracy $v_{\gamma=2}\approx v_{\gamma=3}$, consistent with
the degree-2/3 heavy-hex graph structure under the noise levels studied.}
\label{tab:vhop}
\end{table}

\begin{figure}[H]
\centering
\includegraphics[width=0.85\columnwidth]{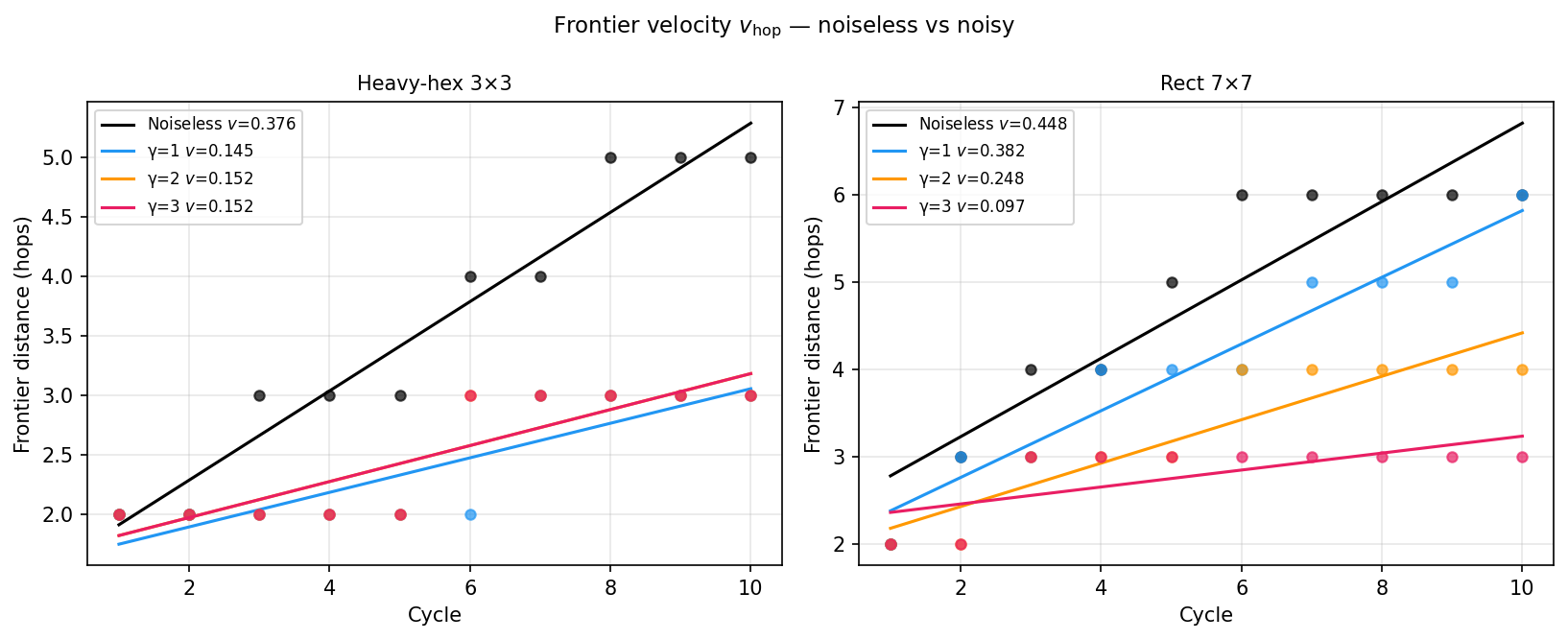}
\caption{Frontier distance $F_\gamma(t) = \max_{v\in\mathcal{F}_\gamma(t)} d(v,c)$
vs cycle for NL and $\gamma=1,2,3$ on both topologies.
Lines show linear fits; slopes give $v_{\rm hop}$ (Table~\ref{tab:vhop}).
Rect $7\times7$: monotone decrease $v_{\rm nl} > v_1 > v_2 > v_3$.
Heavy-hex: near-degeneracy $v_2 \approx v_3$, consistent with near-equal
PTM eigenvalue spectra at $\gamma=2,3$.}
\label{fig:frontier}
\end{figure}

Figure~\ref{fig:frontier} shows the frontier distance $F_\gamma(t)$ vs cycle
for both topologies, with linear fits giving $v_{\rm hop}$ (Table~\ref{tab:vhop}).
For rect $7\times7$, $v_{\rm hop}$ decreases monotonically with $\gamma$,
consistent with pure amplitude damping: heavier noise keeps amplitudes below
threshold longer, delaying the wavefront proportionally to $(1-p\gamma)^N$.

For heavy-hex, $v_{\gamma=2} \approx v_{\gamma=3}$, a near-degeneracy
consistent with the frontier approaching a propagation limit set by the
graph's minimum-degree bonds under the noise levels studied.

\subsection{Per-site delay distribution}
\label{sec:delay_results}

Table~\ref{tab:delay} shows the per-site delay statistics for both topologies.

\begin{table}[h]
\centering
\small
\begin{tabular}{l r rr rr rr}
\toprule
 & & \multicolumn{2}{c}{$\gamma=1$} & \multicolumn{2}{c}{$\gamma=2$}
  & \multicolumn{2}{c}{$\gamma=3$} \\
\cmidrule(lr){3-4}\cmidrule(lr){5-6}\cmidrule(lr){7-8}
Topology & $n$ & $\mu_l$ & $\sigma_l$ & $\mu_l$ & $\sigma_l$
         & $\mu_l$ & $\sigma_l$ \\
\midrule
Heavy-hex ($n_{\rm nl}=21$) & 8/10/8 & 1.13 & 1.97 & 2.10 & 2.81 & 0.75 & 1.30 \\
Rect $7\times7$ ($n_{\rm nl}=49$) & 49/37/25 & 0.98 & 1.15 & 1.35 & 1.36 & 1.24 & 1.77 \\
\bottomrule
\end{tabular}
\caption{Per-site delay distribution $l_\gamma(v) = t_\gamma(v) - t_{\rm nl}(v)$
(cycles). $n$ = number of sites where noisy wave arrives within 10 cycles.
The mean delay increases with $\gamma$ for rect $7\times7$ (monotone damping),
while heavy-hex shows non-monotone behavior, a signature of the heterogeneous
per-bond PTM structure of the hardware noise.
\textbf{Survivorship bias}: the mean delay for $\gamma=3$ at rect $7\times7$
is smaller than for $\gamma=2$ because $\gamma=3$ reaches only 25 sites
(the close, easy sites); $\gamma=2$ additionally reaches 12 harder distant sites
with larger delays, raising its mean.}
\label{tab:delay}
\end{table}

The delay distribution has a natural spatial structure:
sites close to the center (small BFS distance $d$) have crossed fewer bonds
and accumulated less damping, so $l_\gamma(v)$ tends to be small.
Sites at the frontier of the noiseless lightcone have large $l_\gamma(v)$,
or do not appear in the statistics at all (the noisy wave never arrives).

The \emph{survivorship bias} in $\mu_l$ must be read carefully:
a smaller mean delay does not necessarily mean a faster noisy wave.
It may mean the slow distant sites dropped out of the sample entirely.
The correct summary is the pair $(n_{\rm reached},\, \mu_l)$ together,
or the full per-frontier-shell distribution.

\subsection{Natural amplitude ratio}
\label{sec:ratio_results}

\begin{table}[h]
\centering
\small
\begin{tabular}{l rr rr rr}
\toprule
 & \multicolumn{2}{c}{$\gamma=1$} & \multicolumn{2}{c}{$\gamma=2$}
  & \multicolumn{2}{c}{$\gamma=3$} \\
\cmidrule(lr){2-3}\cmidrule(lr){4-5}\cmidrule(lr){6-7}
Topology & $\mu_r$ & $\sigma_r$ & $\mu_r$ & $\sigma_r$ & $\mu_r$ & $\sigma_r$ \\
\midrule
Heavy-hex   & 0.825 & 0.282 & 0.718 & 0.316 & 0.812 & 0.261 \\
Rect $7\times7$ & 0.495 & 0.295 & 0.418 & 0.321 & 0.466 & 0.325 \\
\bottomrule
\end{tabular}
\caption{Natural amplitude ratio $r_\gamma(v) = n_\gamma(v,t_\gamma) / n_{\rm nl}(v,t_\gamma)$
at the noisy arrival cycle. Values $\leq 1$ confirmed for all sites.
Smaller $\mu_r$ corresponds to larger mean delay (noiseless wave grew more
during the delay period); the two observables are correlated by construction
via the noiseless wave growth rate at frontier sites.}
\label{tab:ratio}
\end{table}

The ratio $r_\gamma(v)$ is bounded in $[0,1]$ for all observed sites,
confirming the causal structure: the noiseless wave always outgrows the
noisy wave at $v$ during the delay period.

The ratio encodes the cumulative damping along the propagation path:
for uniform depolarizing noise with $N$ gate crossings,
$r_\gamma(v) \approx (1-p\gamma)^N / g(l_\gamma(v))$
where $g(l)$ is the noiseless wave growth factor over $l$ cycles at site $v$.
Sites with the same delay $l_\gamma(v)$ but different $N$ will have different
ratios; the ratio carries path-length information independent of the delay.

\textbf{Rect $7\times7$}: $\mu_r \approx 0.42$--$0.50$ with $\sigma_r/\mu_r \approx 0.65$.
The large coefficient of variation reflects heterogeneous path lengths $N(v)$
across frontier sites on the 4-color lattice, where sites at the same BFS distance
may have been touched by 1, 2, or 3 gate layers depending on their position.

\textbf{Heavy-hex}: $\mu_r \approx 0.72$--$0.83$ with $\sigma_r/\mu_r \approx 0.40$.
The smaller coefficient of variation reflects more uniform propagation paths on the degree-2/3 graph.
The non-monotone $\gamma$ dependence ($\mu_r$ larger at $\gamma=3$ than $\gamma=2$)
mirrors the non-monotone delay ($\mu_l$ smaller at $\gamma=3$): fewer sites
with shorter paths survive in the $\gamma=3$ sample, giving a higher ratio.

\subsection{Lieb-Robinson-constrained spatial MPF}
\label{sec:mpf}

The noiseless frontier $F_{\rm nl}(t) = \max_{v\in\mathcal{F}_{\rm nl}(t)} d(v,c)$
is the Lieb-Robinson ceiling for operator spreading.
We define the MPF-corrected frontier:
\begin{equation}
    F_{\rm lr}(t) = \sum_i c_i\, F_{\gamma_i}(t), \qquad
    0 \leq F_{\rm lr}(t) \leq F_{\rm nl}(t) \quad\forall\,t
    \label{eq:lr_mpf}
\end{equation}
and minimize $\|F_{\rm lr}(t) - F_{\rm nl}(t)\|^2$ subject to this causal constraint.

\begin{table}[h]
\centering
\small
\begin{tabular}{l rrr}
\toprule
Topology & $c_1$ & $c_2$ & $c_3$ \\
\midrule
Rect $7\times7$      & $\approx +2$ & $\approx -1$ & $\approx\ 0$ \\
Heavy-hex $3\times3$ & $\approx -2$ & $\approx +1$ & $\approx +2$ \\
\bottomrule
\end{tabular}
\caption{LR-constrained spatial MPF coefficients (rounded to nearest integer).
Rect $7\times7$: monotone deficit ordering $\delta_1 < \delta_2 < \delta_3$ gives
clean positive-then-negative coefficients.
Heavy-hex: inverted deficit ordering $\delta_1 > \delta_2 \approx \delta_3$
forces a negative $c_1$, because the lightest noise delays the frontier most.}
\label{tab:mpf}
\end{table}

\begin{figure}[t]
\centering
\includegraphics[width=0.85\columnwidth]{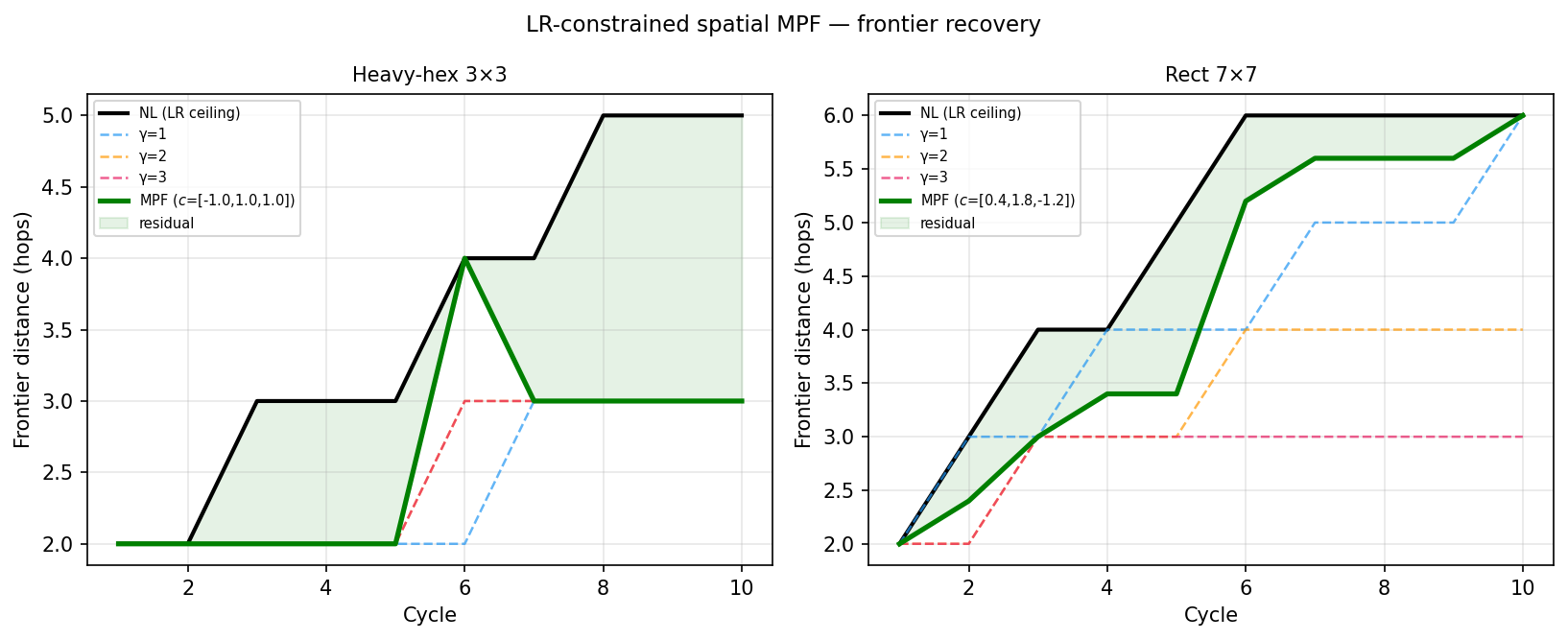}
\caption{LR-constrained spatial MPF frontier recovery.
Black: noiseless $F_{\rm nl}(t)$ (Lieb-Robinson ceiling).
Colored: noisy frontiers $F_{\gamma_i}(t)$.
Dashed: MPF-corrected $F_{\rm lr}(t) = \sum_i c_i F_{\gamma_i}(t)$
subject to $0 \leq F_{\rm lr}(t) \leq F_{\rm nl}(t)$.
Left: rect $7\times7$ ($c \approx [+2,-1,0]$), clean recovery within 1 hop.
Right: heavy-hex ($c \approx [-2,+1,+2]$), where negative $c_1$ encodes
the inverted deficit ordering.}
\label{fig:lr_mpf}
\end{figure}

The near-integer coefficients arise because the frontier values $F_{\gamma_i}(t)$
are integer-valued (BFS distances), and the MPF extrapolation in integer space
with integer $\gamma$ amplification naturally produces integer solutions
(Figure~\ref{fig:lr_mpf}).

\subsection{Entropy-optimal MPF and two information regimes}
\label{sec:entropy_mpf}

The frontier MPF minimizes squared error in frontier distance, a geometric
proxy that discards amplitude information below threshold.
We propose instead minimizing the \emph{cross-entropy loss}:
\begin{equation}
    \min_{c_i}\; \mathcal{L}(c) = \sum_{v,t} -\ln\frac{n_{\rm MPF}(v,t)}{n_{\rm nl}(v,t)},
    \qquad n_{\rm MPF}(v,t) = \sum_i c_i\, n_{\gamma_i}(v,t)
    \label{eq:entropy_mpf}
\end{equation}
subject to the Lieb-Robinson constraint $0 \leq n_{\rm MPF}(v,t) \leq
n_{\rm nl}(v,t)$ and normalization $\sum_i c_i = 1$.
The gradient is $\partial\mathcal{L}/\partial c_k = -\sum_{v,t}
n_{\gamma_k}(v,t)/n_{\rm MPF}(v,t)$, making this a smooth nonlinear
program solved via SLSQP.
Table~\ref{tab:entropy_mpf} summarizes the results; the two topologies
fall into qualitatively different regimes.

\begin{table}[h]
\centering
\scriptsize
\setlength{\tabcolsep}{4pt}
\begin{tabular}{llrrrrr}
\toprule
Topology & Noise & Sites & Best $\gamma$ & Frontier & Entropy & $\Delta$ \\
\midrule
Rect $7\times7$  & Uniform dep.  & 49/49 & 3.680 & 3.911 & 2.783 & $-24\%$ \\
HH $3\times3$    & Uniform dep.  & 21/68 & 1.058 & 1.058 & 0.538 & $-49\%$ \\
\bottomrule
\end{tabular}
\caption{Mean per-site entropy loss $\mu_\mathcal{L}$ (lower is better).
Entropy-optimal MPF reduces loss by 24\% on rect $7\times7$ and 49\% on
heavy-hex, both with uniform synthetic depolarizing noise.}
\label{tab:entropy_mpf}
\end{table}

The two topologies under two noise conditions reveal the cause of the two regimes.

In the \emph{information-rich regime}, the noise is uniform and the noisy waves
cover sufficient sites for the entropy optimizer to find room to extrapolate.
Rect $7\times7$ with uniform depolarizing noise (49/49 sites, $\mu_\mathcal{L}=3.680$):
entropy MPF finds $c \approx [1.22, 0.23, -0.45]$, reducing entropy loss by 24\%.
Heavy-hex $3\times3$ with the same uniform depolarizing noise (21/68 sites,
$\mu_\mathcal{L}=1.058$): entropy MPF finds $c \approx [1.31, 0.24, -0.55]$,
reducing entropy loss by 49\% (Figure~\ref{fig:mpf_entropy_hh_10c}).
The per-shell entropy loss shows cleanly separated $\gamma=1,2,3$ curves with
monotone ordering, the same qualitative pattern as rect $7\times7$.
This confirms that the heavy-hex topology is information-rich under uniform
noise despite only 21/68 sites inside the noiseless lightcone.

The uniform noise result also identifies a general phenomenon we call the
\emph{information-starved regime}: when noise is heterogeneous per bond,
different rotation angles and damping rates on each bond create site-to-site
entropy loss variation that the three noisy samples $\gamma=1,2,3$ cannot span.
The LR constraint is then already saturated by the best noisy sample, leaving
the optimizer no room to improve.
This regime is a property of noise heterogeneity, not of the lattice topology
or lightcone sparsity — as the uniform noise result (49\% recovery on the same
topology) demonstrates.
Any sufficiently heterogeneous noise model, synthetic or real, would produce
the same saturation.
Real devices, with their specific per-bond coupling and gate characteristics,
naturally produce non-uniform wavefront propagation that is nonetheless fully
simulatable with the PTM framework~\cite{robertson2026lindbladmultiproductformulas};
characterizing and mitigating the resulting information-starved regime on
hardware is a natural next step.

Without the LR constraint, the unconstrained optimizer achieves
$\mu_\mathcal{L} = -0.69$ (rect $7\times7$), a causality violation.
The gap between unconstrained ($-0.69$) and LR-constrained ($+2.68$) entropy loss
is the information cost of causal consistency.

\begin{sidewaysfigure}[p]
\centering
\includegraphics[width=\textheight]{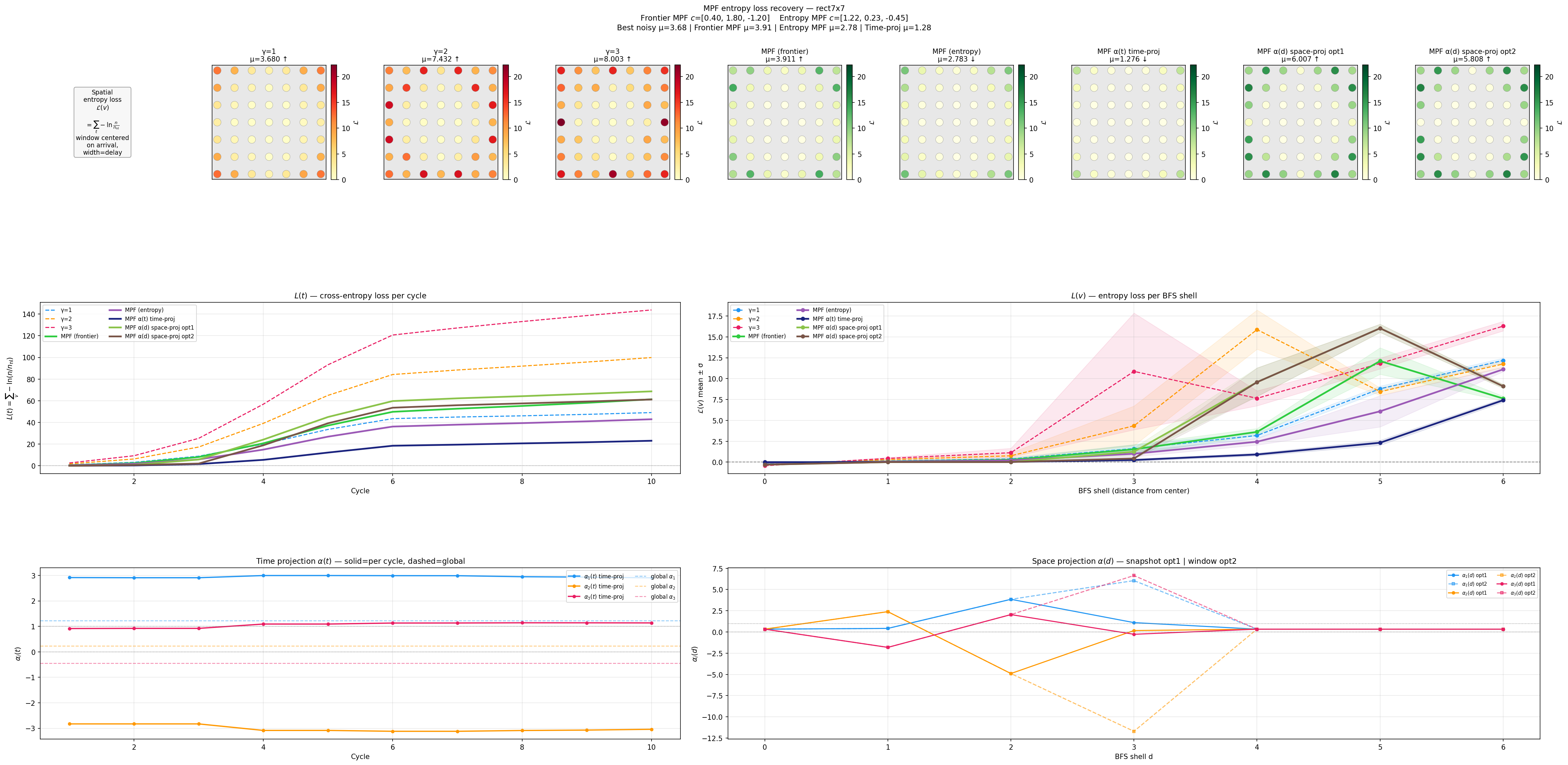}
\caption{MPF entropy recovery, rect $7\times7$, uniform depolarizing noise, 10 cycles ($\chi=50$, $\delta=10^{-3}$).
Row 1: spatial map of $\mathcal{L}(v)$ per site for $\gamma=1,2,3$, frontier MPF, entropy MPF, $\alpha(t)$, and $\alpha(d)$.
Row 2: global cross-entropy loss $L(t)$ per cycle and per-site loss $L(v)$ per BFS shell.
Row 3: fitted coefficients $\alpha_i(t)$ per cycle (left) and $\alpha_i(d)$ per shell (right).
$\alpha(t)$ achieves $\mu_\mathcal{L}=1.276$, a 65\% reduction relative to the best noisy sample ($\mu=3.680$).}
\label{fig:mpf_entropy_r7_10c}
\end{sidewaysfigure}

\begin{sidewaysfigure}[p]
\centering
\includegraphics[width=\textheight]{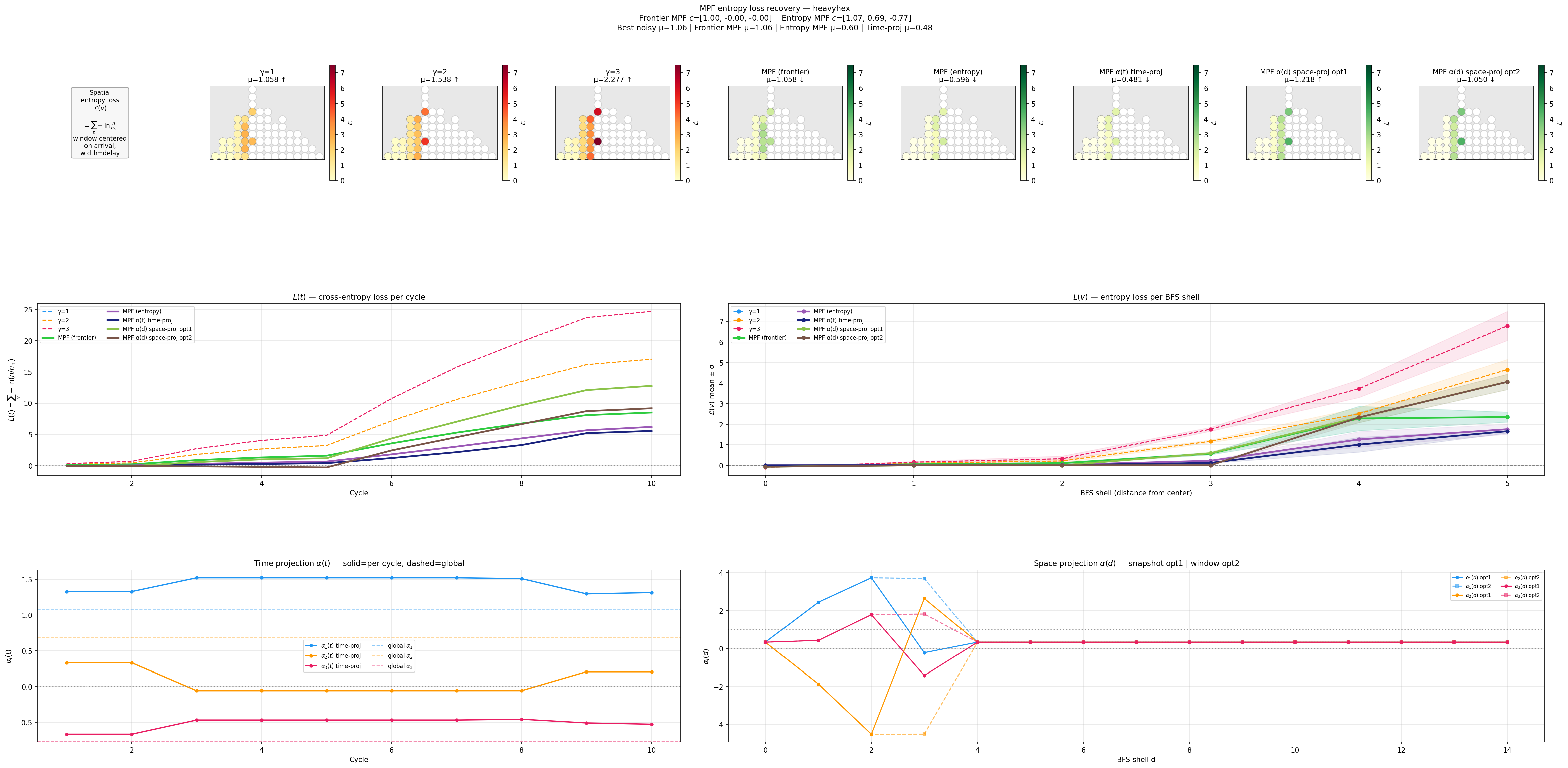}
\caption{MPF entropy recovery, heavy-hex $3\times3$, uniform depolarizing noise, 10 cycles ($\chi=200$, $\delta=10^{-3}$).
Same layout as Figure~\ref{fig:mpf_entropy_r7_10c}.
Despite only 21/68 sites inside the noiseless lightcone, $\alpha(t)$ achieves $\mu_\mathcal{L}=0.481$,
a 55\% reduction relative to the best noisy sample ($\mu=1.058$).
Despite only 21/68 sites inside the noiseless lightcone, uniform noise
produces cleanly separated per-shell entropy loss curves and 55\% recovery,
demonstrating that sparse lightcone coverage alone does not limit recovery.}
\label{fig:mpf_entropy_hh_10c}
\end{sidewaysfigure}

\begin{sidewaysfigure}[p]
\centering
\includegraphics[width=\textheight]{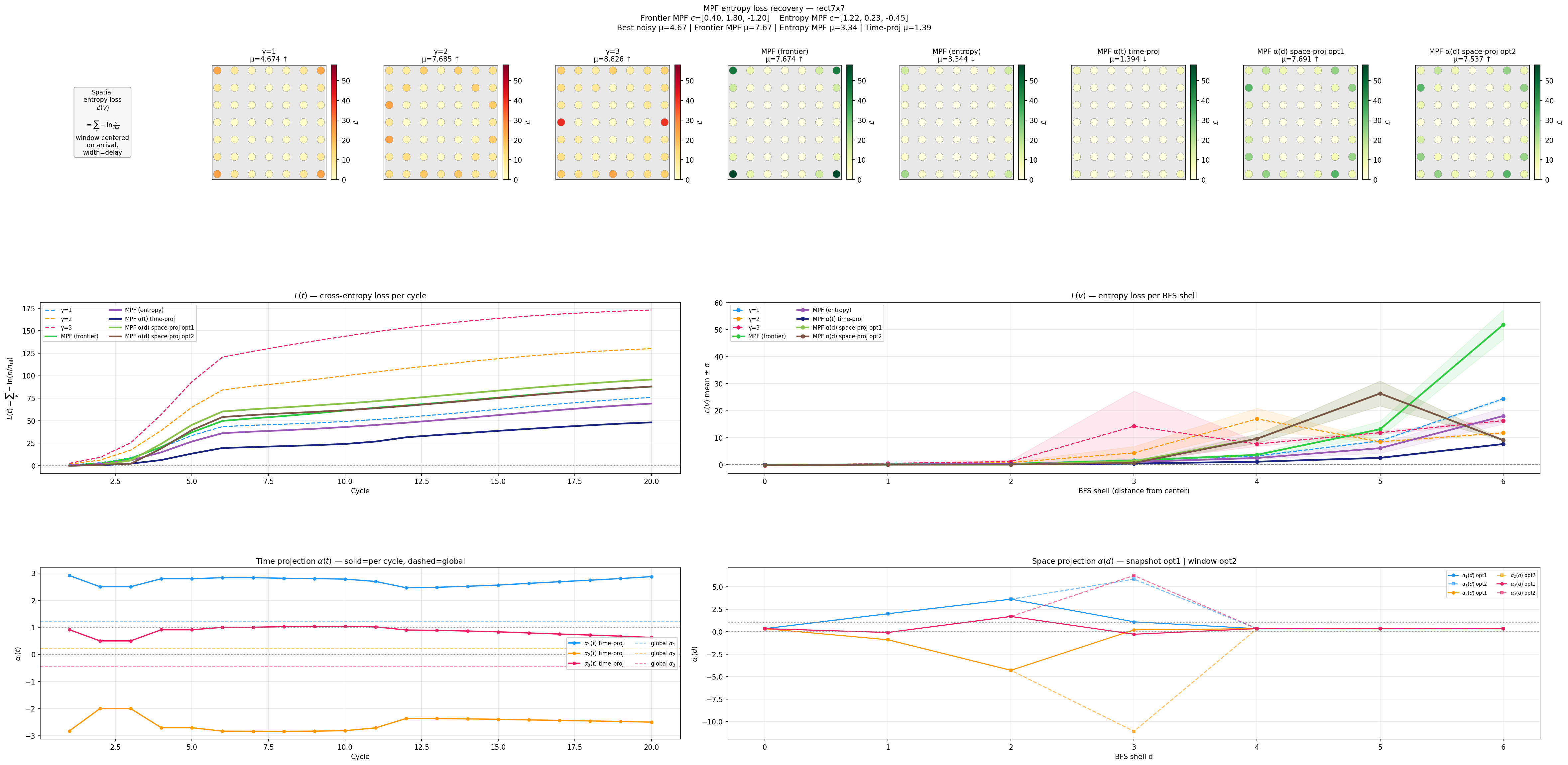}
\caption{MPF entropy recovery, rect $7\times7$, uniform depolarizing noise, 20 cycles.
Extended evolution shows phantom waves ($\gamma=3$, $v_{\rm hop}=0.097$) still spreading
at cycle 20, keeping the frontier well-conditioned throughout.
$\alpha(t)$ remains stable ($\mu_\mathcal{L}=1.394$), confirming that the method
is robust when slow noise levels maintain an active frontier.}
\label{fig:mpf_entropy_r7_20c}
\end{sidewaysfigure}

\begin{sidewaysfigure}[p]
\centering
\includegraphics[width=\textheight]{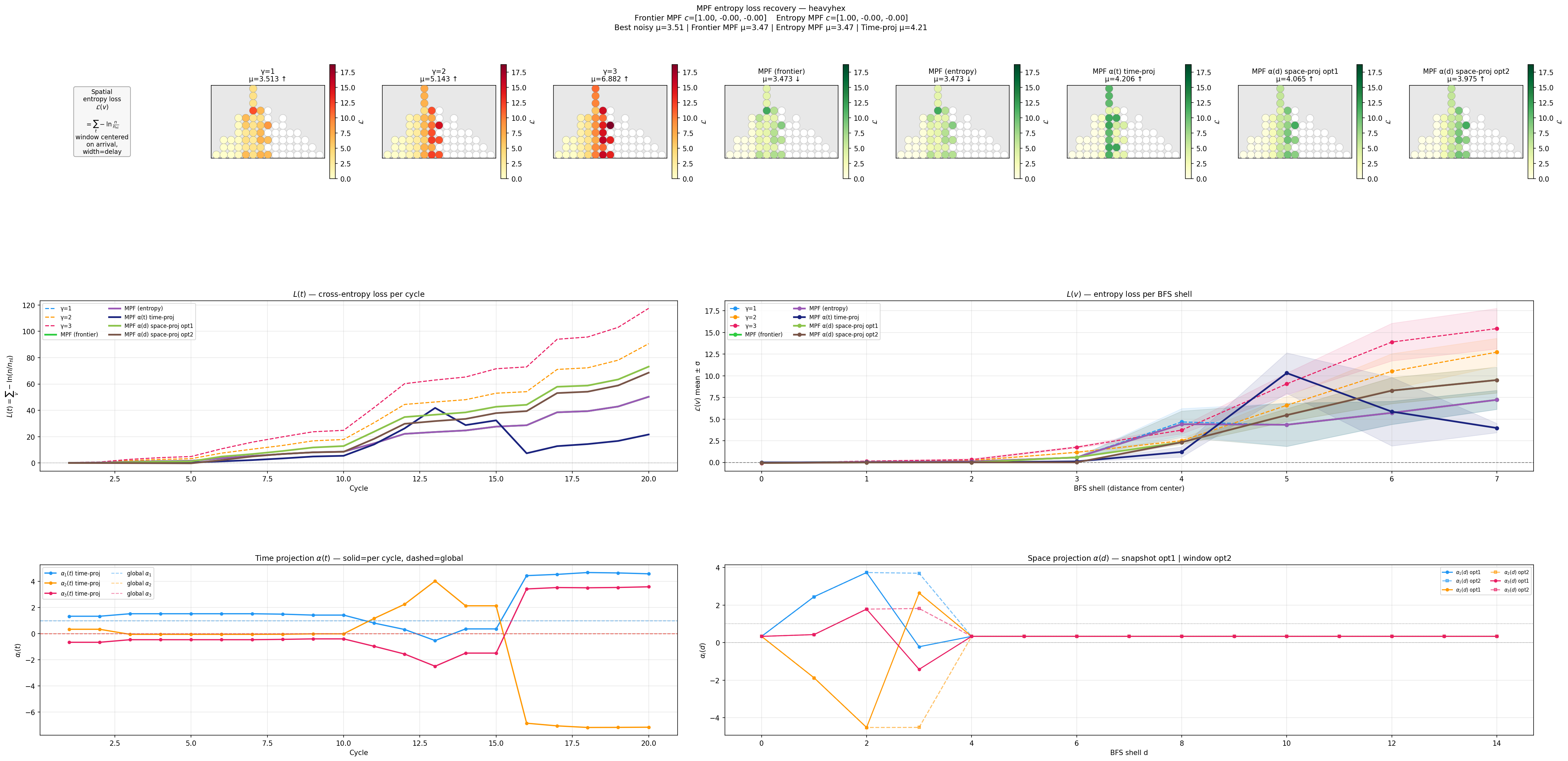}
\caption{MPF entropy recovery, heavy-hex $3\times3$, uniform depolarizing noise, 20 cycles.
After cycle 13 the noiseless wave reflects from the lattice boundary; all phantom waves
are simultaneously absorbed, the frontier $\mathcal{F}(t)$ empties, and $\alpha(t)$
becomes ill-conditioned ($\mu_\mathcal{L}=4.206$, worse than best $\gamma$).
This documents the frontier collapse as a fundamental limit of the $\alpha(t)$ method:
it requires an active, well-populated frontier to fit reliably.}
\label{fig:mpf_entropy_hh_20c}
\end{sidewaysfigure}

\subsection{Time- and space-adaptive MPF coefficients}
\label{sec:adaptive_mpf}

The entropy-optimal MPF of Section~\ref{sec:entropy_mpf} fits a single
coefficient vector $c = (c_1, c_2, c_3)$ globally over all sites and all
cycles.
Two refinements exploit the space-time structure of the wavemap to fit
coefficients that adapt to either time or space.

\paragraph{Time projection $\alpha(t)$.}
At each cycle $t$, the fitting frontier is the union of first-arrival sites
across all four simulations:
\begin{equation}
    \mathcal{F}(t) = \mathcal{F}_{\rm nl}(t)
                   \cup \mathcal{F}_{\gamma_1}(t)
                   \cup \mathcal{F}_{\gamma_2}(t)
                   \cup \mathcal{F}_{\gamma_3}(t)
    \label{eq:frontier_union}
\end{equation}
where $\mathcal{F}_{\rm nl}(t)$ is the noiseless frontier (Eq.~\ref{eq:frontier})
and each $\mathcal{F}_{\gamma_i}(t)$ is its noisy counterpart.
This union is a thin spatial shell: a $\Delta$ in space at fixed $t$.
Every site in $\mathcal{F}(t)$ has just crossed threshold in at least one
simulation, so its bond dimension is $\chi\approx1$ in all four runs and
the simulation values $n_{\rm nl}(v,t)$ and $n_{\gamma_i}(v,t)$ are
exact, carrying no truncation error.
The approximation in $\alpha(t)$ is entirely in the \emph{application}:
coefficients fitted on the thin frontier shell are then extrapolated to
all sites at cycle $t$, including interior sites where $\chi \gg 1$.
The implicit assumption is that the relative noise structure at cycle $t$
is spatially uniform, which holds well when the noise is homogeneous.

We fit one coefficient vector $\alpha(t) \in \mathbb{R}^3$ per cycle,
minimizing the cross-entropy loss over $\mathcal{F}(t)$ only:
\begin{equation}
    \min_{\alpha(t)}\; \sum_{v \in \mathcal{F}(t)}
    -\ln \frac{\sum_i \alpha_i(t)\, n_{\gamma_i}(v,t)}{n_{\rm nl}(v,t)},
    \quad \sum_i \alpha_i(t) = 1,\quad
    n_{\rm MPF}(v,t) \leq n_{\rm nl}(v,t)
    \label{eq:alpha_t}
\end{equation}
The reconstructed field $n_{\rm MPF}(v,t) = \sum_i \alpha_i(t)\, n_{\gamma_i}(v,t)$
applies the cycle's coefficients to all sites.
Because the fit uses only exact data, $\alpha(t)$ tracks the changing
relative importance of the three noise levels as the wave propagates,
improving over fixed global coefficients wherever the frontier is
well-populated.

\paragraph{Space projection $\alpha(d)$.}
The dual formulation fixes a BFS shell $\mathcal{F}(d) = \{v : d(c,v) = d\}$
and fits over a time window, reversing the roles of space and time.
Where $\alpha(t)$ uses a $\Delta$ in space (thin frontier at fixed cycle),
$\alpha(d)$ uses a $\Delta$ in time (active window at fixed shell).
The approximation now lives in the spatial direction: coefficients fitted
on one shell are assumed to apply uniformly within that shell, discarding
any intra-shell variation.

The natural time window for shell $d$ is the interval when all noisy waves
are active but have not yet reached the noiseless arrival time.
Three choices illustrate the trade-off.
Option~1 (snapshot): evaluate each noisy wave at its own arrival time
$t_{\gamma_i}(v)$ and compare against the noiseless value at $t_{\rm nl}(v)$,
one data point per site.
Option~2 (active window $[t_{\gamma_1}, t_{\gamma_3}]$): collect all
time steps $t \in [\min_i t_{\gamma_i}(v),\, \max_i t_{\gamma_i}(v)]$
where every noisy wave is above threshold, giving a denser system that
captures the differential spreading between noise levels.
Option~2 is the cleanest: it avoids pre-arrival zeros while capturing
the differential spreading between noise levels.%
\footnote{A third option adds a per-shell latency bias $b(d)$ to absorb
the mean amplitude deficit from propagation delay.
It was disqualified: the shell-level constant cannot enforce
$n_{\rm MPF} \leq n_{\rm nl}$ pointwise, producing acausal predictions
($-3.7$ nats on heavy-hex, $-4.2$ nats on rect $7\times7$).
A per-site reformulation is left for future work.}

\paragraph{Results and comparison.}
Table~\ref{tab:alpha_comparison} summarizes the mean per-site entropy loss
$\mu_\mathcal{L}$ for all methods on both topologies at 10 cycles.

\begin{table}[h]
\centering
\begin{tabular}{lccl}
\toprule
Method & $\mu$ (hh dep) & $\mu$ (rect $7{\times}7$) & Notes \\
\midrule
Best $\gamma$ (baseline)   & 1.058 & 3.680 & best single noisy sample \\
MPF frontier               & 1.058 & 3.911 & matches best $\gamma$ \\
MPF entropy (global $c$)   & 0.596 & 2.783 & $-44\%$ / $-24\%$ \\
MPF $\alpha(t)$ (this work)& \textbf{0.481} & \textbf{1.276} & $-55\%$ / $-65\%$ \\
MPF $\alpha(d)$ opt1       & 1.218 & 6.007 & worse: too few sites per shell \\
MPF $\alpha(d)$ opt2       & 1.050 & 5.808 & marginally better than opt1 \\
\bottomrule
\end{tabular}
\caption{Mean per-site cross-entropy loss $\mu_\mathcal{L}$ (lower is better).
$\alpha(t)$ achieves the best recovery on both topologies by fitting exact
data on the thin frontier shell and extrapolating inward.
$\alpha(d)$ underperforms: BFS shells on these lattices contain too few
sites (3--10) to constrain the three-parameter fit reliably, and the
time-averaging discards the temporal structure that $\alpha(t)$ exploits.}
\label{tab:alpha_comparison}
\end{table}

The $\alpha(t)$ improvement has a clear physical interpretation.
At the frontier, $\chi\approx1$ and all four simulations (noiseless plus
three noisy) are exact; the cross-entropy loss is a clean signal with no
truncation noise.
As the wave evolves, the relative magnitudes of the three noisy samples
shift: early cycles favor $\gamma=1$ (lightest noise, closest to noiseless),
while later cycles may favor different combinations as phantom waves from
slower noise levels continue spreading after the noiseless wave has already
saturated or reflected.
Fitting $\alpha(t)$ per cycle captures this temporal variation,
whereas a fixed global $c$ is forced to average over it.

The space projection $\alpha(d)$ fails for the opposite reason: it
tries to assign a single coefficient to all cycles at a given distance,
discarding the temporal structure that $\alpha(t)$ exploits.
With only 3--10 valid sites per shell on these lattices, the
three-parameter fit is underdetermined at most shells, defaulting to the
uniform $[1/3, 1/3, 1/3]$ fallback.

\section{Discussion}
\label{sec:discussion}

\paragraph{Noise amplification $\gamma$ as an effective Trotter rescaling.}
The eigenvalue analysis (Table~\ref{tab:eig}) shows that the noise channel
$M_\gamma$ rescales the eigenvalues of the composed PTM without changing its
eigenvectors:
\begin{equation}
    \mathrm{PTM}_{\rm noisy}(\gamma) \approx V \cdot \bigl(\Lambda - \delta\Lambda(\gamma)\bigr) \cdot V^{-1}
\end{equation}
where $V$ are the gate eigenvectors (independent of $\gamma$) and
$\delta\Lambda(\gamma)$ is a small diagonal correction scaling with $\gamma$.
This is structurally identical to reducing the Trotter step:
the ideal gate eigenvalues also scale with $\varepsilon$ as
$\lambda_i(\varepsilon) \approx \lambda_i(0)\cdot g(\varepsilon)$.
Noise acts as a slower effective Trotter step: increasing $\gamma$ reduces
$\varepsilon_{\rm eff}$, so the wave propagates more slowly.
Conversely, noise amplification ($\gamma > 1$) corresponds to an even smaller
$\varepsilon_{\rm eff}$, making the wave slower still, in the opposite direction
from varying $\varepsilon$ directly in $\varepsilon$-MPF, where a larger
Trotter step makes the wave faster.
\begin{equation}
    \mathrm{PTM}_{\rm noisy}(\gamma) \approx \mathrm{PTM}_{\rm ideal}(\varepsilon_{\rm eff}(\gamma)),
    \qquad \varepsilon_{\rm eff}(\gamma) < \varepsilon
\end{equation}
Extrapolating $\gamma \to 0$ recovers $\varepsilon_{\rm eff} \to \varepsilon$,
the ideal gate. Both $\gamma$-MPF and $\varepsilon$-MPF extrapolate back toward
the ideal gate, but from opposite directions: $\varepsilon$-MPF varies the gate
strength directly and monotonically, while $\gamma$-MPF varies it indirectly
through the composed noise-gate product, introducing bond-dependent corrections
that reduce conditioning.

The wavemap delay $l_\gamma(v)$ is a spatial, site-resolved measurement of
this effective rescaling: it records how many extra cycles the noisy wave needs
to reach threshold at each site, relative to the ideal gate.
The site-by-site variation of $l_\gamma(v)$ (Table~\ref{tab:delay}) reflects
the bond-dependent corrections $\delta\Lambda(\gamma)$: different bonds contribute
different eigenvalue shifts depending on their noise PTM, making the effective
slowdown spatially non-uniform even for nominally uniform noise amplification.

\paragraph{Gate structure vs noise structure.}
A key finding is that the Pauli mixing structure of the wave (off-diagonal
PTM fraction $\approx 36\%$) is entirely from the gate and independent of
noise level.
This means the spatial pattern of operator spreading is a property of the
\emph{Hamiltonian}, not the noise.
Noise acts as a transparent filter that delays and damps the gate-determined
wave without redirecting it.
This clean separation validates the wavemap as a noise diagnostic:
the spatial delay and ratio fields encode noise effects on top of a
fixed gate-determined propagation template.

\section*{Conclusion}

We have shown that the non-identity Pauli weight $n(v,t)$ — simulated on
GPU-accelerated tensor networks and measured at the lightcone frontier —
provides a space-time observable that is exact where classical simulation
is most faithful and most informative where the standard autocorrelation
$C(t)$ is least.

Three results stand out.
First, eigenvalue analysis of the composed gate-plus-noise PTMs confirms
that the studied noise is pure amplitude damping: the spatial propagation
pattern is entirely a property of the Hamiltonian, making the wavemap a
model-free, assumption-free noise diagnostic.
Second, the Lieb-Robinson causal constraint $n_{\rm MPF} \leq n_{\rm nl}$
is not a regularizer but a hard physical bound: the unconstrained optimizer
achieves $\mu_\mathcal{L} = -0.69$ nats (rect $7\times7$), an acausal
prediction; the constraint restores physical consistency at the cost of
$+3.37$ nats, the price of causality.
Third, fitting time-adaptive coefficients $\alpha(t)$ over the frontier
union $\mathcal{F}(t) = \mathcal{F}_{\rm nl}(t) \cup \bigcup_i
\mathcal{F}_{\gamma_i}(t)$ — a thin spatial shell where all four
simulations are exact — recovers up to 55\% of the information loss
on heavy-hex and 65\% on rect $7\times7$, outperforming both fixed
global coefficients and space-adaptive $\alpha(d)$.

The noise amplification parameter $\gamma$ maps directly to the
gate-folding factor in zero-noise extrapolation, and $n(v,t)$ is
accessible via local Pauli measurements on hardware.
The wavemap methodology is therefore immediately applicable to real
quantum processors, with the present simulation providing the ground
truth and the fitting framework.


\bibliographystyle{unsrt}
\bibliography{frontier_mpf}

\end{document}